\documentclass[8pt]{article}
\usepackage{graphicx}
\usepackage{pgfplots}
\usepackage{newlfont}
\usepackage[colorlinks,linkcolor=blue,citecolor=blue,
anchorcolor=blue,bookmarksopen,pdfpagetransition={Wipe}]{hyperref}
\usepackage{amsmath,amsthm,amssymb}
\usepackage{subfigure}
\usepackage{graphicx}
\usepackage{amsmath, amsfonts}
\usepackage{dsfont}
\usepackage{booktabs}
\usepackage{fourier}
\usepackage{array}
\usepackage{makecell}
\usepackage{algorithm}
\usepackage{algpseudocode}
\usepackage{lscape}
\usepackage{threeparttable}
\usepackage[colorlinks]{hyperref}
\usepackage[T1]{fontenc}
\usepackage[tableposition=top]{caption}
\usepackage{tabularx}
\newtheorem{thm}{Theorem}[section]

\newtheorem{prop}[thm]{Property}

\numberwithin{equation}{section}
\usepackage{float}

\DeclareMathOperator{\sign}{sign}
\hypersetup{
urlcolor=blue	
}
\begin{document}
%--------------------------title-----------------Corresponding author:-------
\title{\textbf{Poissonian Blurred Image Deconvolution by Framelet
based Local Minimal Prior}}
\author{Reza Parvaz\footnote{ Corresponding author: rparvaz@uma.ac.ir, reza.parvaz@yahoo.com}}
\date{}
\maketitle
\begin{center}
	Department of Mathematics, University of Mohaghegh Ardabili,
	56199-11367 Ardabil, Iran.\\
\end{center}
\date{}
\maketitle
%------------------------------------abstract--------------------------------
\begin{abstract}
\noindent
Image production tools do not always create a clear image, noisy and blurry images are sometimes created. Among these cases, Poissonian noise is one of the most famous noises that appear in medical images and images taken in astronomy.
Blurred image with Poissonian noise obscures important details that are of great importance in medicine or astronomy. Therefore, studying and increasing the quality of images that are affected by this type of noise is always considered by researchers.
In this paper, in the first step, based on framelet transform, a local minimal prior is introduced, and in the next step, this tool together with fractional calculation is used for Poissonian blurred image deconvolution. In the following, the model is generalized to the blind case. To evaluate the performance of the presented model, several images such as real images have been investigated.
\end{abstract}
%------------------------------------------Keywords---------------------------
\vskip.3cm

\section{Introduction}
Image is one of the most important tools in the development of various sciences, such as medicine and astronomy. 
In these sciences, the details of the image are very important in diagnosing the disease or analyzing an image. Therefore, removing a detail from the image can cause unfavorable image analysis. Also despite the development of imaging tools in astronomy and medicine, it has not been possible to create a completely clear image.
Noise and blur are two elements that affect image quality.
Deblurring problem is influenced by the point spread function (PSF) and is divided into two types: non-blind and blind problems. In the first type, we have information about the PSF. And in the second case, only the blurred image is available. 
There are different types of PSF in this discussion, such as average,
disk, gaussian, motion. These types of functions are defined in the image processing software such as Matlab or OpenCV and can be easily used.
Like the PSF, there are different types of image noise that reduce image quality, such as the 
Poissonian, salt and pepper and periodic noises. Poissonian noise also called Photon noise
has a Poissonian distribution and appears in medical images as X-ray image \cite{1} and astronomical images \cite{2}.
The reader can find more information about the structure of this type of noise in \cite{3}. 
The general structure of Poissonian blurred image is written as follows
\begin{align}\label{e1}
Y=\mathbf{P}(K\circledast X),
\end{align}
where $Y$ and $X \in \mathbb{R}^{n \times m}$ denote blurred and clear images, respectively,
$K\in \mathbb{R}^{l \times s}$ shows the point spread function and 
 $\circledast$ represents two dimensional convolution operator. 
$\mathbf{P}$ also represents the process of adding Poissonian noise to the blurred image.
This type of problem can be expressed as a system of linear equations as follows
\begin{align}\label{e2}
y=\mathbf{P}(k x),
\end{align}
where $y,x \in \mathbb{R}^{nm \times 1}$  and $k \in \mathbb{R}^{nm \times nm}$.
This system of equations can not be solved directly due to the large volume and ill-conditioned. Also, the structure of the blurred kernel matrix depends on the boundary conditions of the problem as zeros and periodic bounders.
It is well known that this matrix can be obtained by the fast Fourier transform(FFT) under the periodic condition. 
Using the Fourier method to create matrices has the advantage of reducing computational volume and program execution time.
The first attempt to solve this type of problem can be to use the singular value decomposition, although this may not be an efficient method despite Poissonian noise.
One of the efficient methods to solve this type of problem is to use the total variation (TV) method \cite{4}, which is presented by the first article and is used in most articles.
The results for the nonblind and blind types of problems using this method show the efficiency of the TV method in obtaining the appropriate solution for deblurring problem.
With the expansion of the discussion on fractional derivatives, the TV method is developed and various articles are used based on fractional derivatives \cite{5,6}.
Another tool used to improve the TV method is the use of framelet transform \cite{7}. This transform provides a sparse approximation and has been used in various articles as \cite{8,9,10}.
In recent years, various methods have been used for Poissonian blurred image deconvolution.
The frame based algorithm by a nonstationary accelerating alternating direction method
is introduced in \cite{11}, an algorithm based on regularized by a denoiser
constraint and deep image prior is used in \cite{12}, also the non-local total variation is introduced in \cite{13}. 
One of the efficient tools used to solve the blurred image deconvolution problem is the dark
channel metric \cite{14,15}. Although this method has a good solution, but the volume of calculations of this method is high.
To solve this problem, in \cite{16}, an algorithm based on the local minimal pixels over non overlapping patches is introduced.
Due to the fact that in this method all pixels of the image are not evaluated, the computational volume is considerably reduced.
In the proposed algorithm, this method is improved by using framelet theory.
According to the simulated results that are compared in the next section, the difference between blurred and clear images is more visible compared to the method in \cite{15,16}. And this sharp discrepancy caused by the use of framelet provides a good tool for solving the blurred image deconvolution.
In addition to the framelet based local minimal prior,
the fractional derivative based on
the discrete form of the G-L definition is used in the proposed algorithm.
Using this method becomes an efficient tool for approximating the edges of a restored image.
In the real world, we only know the information inside the picture frame.
Lack of data outside the frame causes problems in image deblurring.
One of these problems that appears in the restored image is the ringing artifacts near the image borders. Different methods to solve this problem are studied in different articles.
In \cite{17}, no-boundary convolution by choosing the valid option in 
Matlab's convolution  is proposed. Also, in \cite{18} a method based on
Fourier domain restoration filter and extrapolated image is proposed. 
In this article, the method that is introduced in \cite{18}
is used in the proposed algorithm.
One of the most important advantages of this method is that the bounders of the restored image are not lost.\\

\noindent\textit{Outline:} The organization of this paper is as follows: Motivation and preliminaries
of the paper are studied in Sect. \ref{sec2}. Also in this section,
framelet based local minimal prior which has an important role in the proposed model
is introduced.
In Sect. \ref{sec3}, the proposed model for Poissonian blurred image deconvolution is introduced and a numerical algorithm is proposed to solve this model.
The results of the numerical algorithm for the proposed model is shown in Sect. \ref{sec4}. 
The conclusion of the paper is given in Sect. \ref{sec5}.\\
\noindent \textit{Notations:} 
The notations used are summarized as:
$F(\cdot)$ and $F^{-1}(\cdot)$  denote the fast Fourier transform (FFT)  
and the inverse fast Fourier transform (IFFT),
$\circledast$ and $\langle\cdot , \cdot \rangle$ are used for 
two dimensional convolution operator and inner product, respectively.
And separable Hilbert space is shown by $H$.\\

\section{Motivation and preliminaries}\label{sec2}
In this section, we introduce the tools and ideas used and study them.

\subsection{Framelet theory and fractional derivative}
A countable set as $X$ in a separable Hilbert space $H$
is named a tight frame if for all $f\in H$ the following relation holds
\begin{align*}
f=\sum_{\psi \in X}\langle f, \psi\rangle \psi.
\end{align*}
The collection $X(\Psi)=\{2^{j/2}\psi_i(2^j.-t); 1 \leq i\leq n,~t,j\in \mathbb{Z} \}$
is called wavelet system for $\Psi=\{ \psi_0,\cdots,\psi_n\}$. 
If the  wavelet system be a tight frame, then this system is called a tight wavelet frame
and each element of this system is named a
framelet. For a compactly supported scaling
function like $\phi$ a wavelet tight frame is obtained by
\begin{align*}
	F\big(\phi_i(2\omega)\big)=F(h_i)F\big(\phi(\omega)\big),~~i=0,\ldots,n.
\end{align*}
The piecewise linear B-spline
framelets with low-pass
and high-pass filters as
$h_0=\frac{1}{4} [1, 2,1],~h_1=\frac{\sqrt{2}}{4}
[1, 0,-1]$ and $h_2=\frac{1}{4}
[-1, 2,-1]$, respectively, is used in proposed model.
It is suggested that the reader find more information about this concept from \cite{7,19}.\\
The concept of derivative can be considered as the most widely used tools in the image processing. With the development of the concept of ordinary derivative to fractional derivatives, the use of this tool has been used in various fields of image processing such as image denoising and image deblurring \cite{20,21}.
Despite having a fixed definition for a ordinary derivative,
various definitions have been developed for the fractional derivative as
Riemann-Liouville, Caputo, Caputo-Fabrizio
and Gr\"{u}nwald-Letnikov (G-L) \cite{22,23}.
Due to the discrete structure of G-L definition, this type of definition is most commonly used in image processing. In the proposed algorithm, this definition is used. 
For a real function $u:\Omega \subseteq \mathbb{R}^2 \rightarrow \mathbb{R}^2$, the fractional gradient of order $\alpha \in \mathbb{R}$
is defined as
\begin{align*}
\nabla^{\alpha} u:=\big(\nabla^{\alpha}_h u,\nabla^{\alpha}_v u\big),
\end{align*}
where
\begin{align*}
&\nabla^{\alpha}_h u(i,j)=\sum^{L-1}_{l=0}\frac{(-1)^l \Gamma(\alpha+1)}{\Gamma(l+1)\Gamma(\alpha-l+1)} u(i-l,j),\\
&\nabla^{\alpha}_v u(i,j)=\sum^{L-1}_{l=0}\frac{(-1)^l \Gamma(\alpha+1)}{\Gamma(l+1)\Gamma(\alpha-l+1)} u(i,j-l).
\end{align*}
\subsection{Framelet based local minimal prior}
\begin{figure}[h!]
	\centering
	\makebox[0pt]{
		\subfigure[clear image($I$)]{\label{fig:gull}\includegraphics[width=1\textwidth]{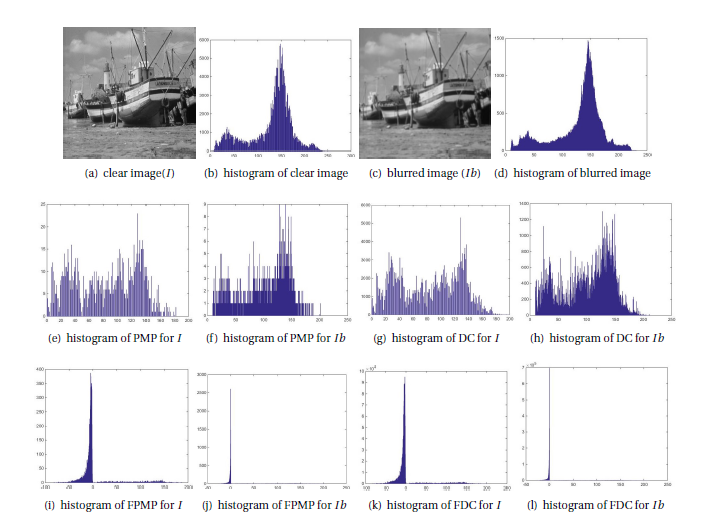}}}\\
	\vspace{-.2cm}
	\caption{Histogram plots for FPMP, FCD, PMP and DC
		of clear and blurred images.}
	\label{fig1}
\end{figure}

For an image $I\in \mathbb{R}^{m\times n \times c}$, the framelet patch-wise minimal pixels (FPMP) is defined as a vector of local minimal pixels over nonoverlapping
patches on the framelet transform and obtained by using
following formula
\begin{align*}
P_{w}(I)(i)=\min_{(x,y)\in  \Omega_i}\big(\min_{c \in {\{r,g,b\}}}WI(x,y,c)\big),
\end{align*}
where $\Omega_i$ means the $i$-th patch of image.
In the above formula, after the framelet transform, the collection of local minimal pixels are calculated for each of the low-pass and high-pass filters.
Therefore,
 if the size of patch is considered
as $r\times r$ then $P_{w}(I)\in\mathbb{R}^{9\varpi}$ where $\varpi=\lceil \frac{m}{r} \rceil.\lceil \frac{n}{r}\rceil$.
Similarly, the framelet dark channel (FDC) is defined as
\begin{align*}
D_{w}(I)(i,j)=\min_{(x,y)\in  \Omega_{i,j}}\big(\min_{c \in {\{r,g,b\}}}WI(x,y,c)\big),
\end{align*}
where $\Omega_{i,j}$ is an image
patch centered at $(i,j)$-th pixel. 
By using this definition, we can see that 
$D_{w}(I)\in\mathbb{R}^{3m \times 3n}$.
Using the same process as \cite{15}, the following 
properties are obtained.
\begin{prop}
Let $P_w(B)$ and $P_w(I)$ show the FPMP of the
blurred and clear images, respectively. Then
\begin{align*}
P_w(B)\geq P_w(I).
\end{align*}
\end{prop}
\begin{prop}
Let $D_w(B)$ and $D_w(I)$ show the FDC of the
blurred and clear images, respectively. Then
\begin{align*}
D_w(B)\geq D_w(I).
\end{align*}
\end{prop}
In addition to the above properties, the histogram plots are drawn in Fig.\ref{fig1}, which shows that the number of zero elements in the FPMP and FDC for the blurred image is greater than these values for the clear image.
Also in this figure, a comparison between the FPMP, FDC, PMP and DC is done.
Based on the results, it can be seen that the difference in the number of non-zero elements between the clear image and the blurred image in the FPMP and FDC is larger than  
the difference of these value in the PMP and DC methods.
Therefore, using the $l_0$-norm along with the FPMP and FDC can be more efficient method in
the deblurring problems; remember that the $l_0$-norm counts the nonzero elements.
Most deblurring algorithms are iterative algorithms and the FDC is large in size
then this increases the computational time of the algorithm. 
To avoid this problem the FPMP is used in the proposed algorithm. 
%%%%%%%%%%%%%%%%%%%%%%%%%%%%%%%%%%%%%%%%%%%%%%%%%%%%%%%%%%%%%%%%%%%%%

\section{Poissonian blurred image deconvolution}\label{sec3}
One of the most common ways to find a clear image $x$ is to use the Poisson assumption for
problem \eqref{e2}. 
The Poisson distribution for blurred problems can be written as
\begin{align*}
\mathbf{P}(y|kx)=\prod^{N}_{i=1} \frac{(kx)_i^{y_i}e^{(kx)_i}}{y_i!}.
\end{align*} 
Based on the negative log-likelihood, we have
\begin{align*}
-\log \mathbf{P}(y|kx)=\langle 1, kx \rangle-\big\langle y, \log(kx)\big\rangle+\sum^{N}_{i=1}\log(y_i!)
\end{align*} 
where $1$ represents a vector that all elements are equal to one. The above problem is an ill-conditioned problem,
therefore, special methods are used to solve this type of problem.
In a common method, two main methods, regularization method and maximum 
a posteriori (MAP) are used to solve this type of problem
In many algorithms, the following variational problem is used to estimate clear image $x$.
\begin{align*}
	\min_{x \geq 0}\mu	\big\langle 1, kx-y \log(kx)\big\rangle+\varphi(x),  
\end{align*}
where $\varphi(x)$ and $\mu>0$ denote for regularizer and  regularization parameters, respectively.\\
\subsection{Non-blind deconvolution problem}
The following model is proposed to solve the Poissonian blurred image deconvolution.
\begin{align}\label{e4}
	\min_{x \geq 0}\mu	\big\langle 1, kx-y \log(kx)\big\rangle+\lambda\| P_w(x)\|_0+\|\nabla^{\alpha} x\|_l+\delta_{\mathbb{R}^N_+}(x),  
\end{align}
where $\delta_{\mathbb{R}^N_+}$ denotes the indicator function over $\mathbb{R}^N_+$.
In this model $l$ is considered as $0$ and $1$. Therefore, two different norms for the proposed model are studied in the proposed model. In the continuation of this paper the model based on $l_0$ norm
is named $l_0$-FPMP  and the model based on $l_1$ norm
is called $l_1$-FPMP.  
In order to approximate the answer, the alternating direction method of multipliers (ADMM) is used.
By using of $z$, $v$ and $m$ as auxiliary variables, Eq. \eqref{e4} is reformulated as 
\begin{align*}
&\min_{x \geq 0}\mu	\langle 1, v-y \log v\rangle+\lambda\| P_w(x)\|_0+\|z\|_l+\delta_{\mathbb{R}^N_+}(m),\\
&\text{s.t.}~~v=kx,~z=\nabla^{\alpha} x,~m=x. 
\end{align*}
Then augmented Lagrangian is written as 
 \begin{align}
	L(x,v,z,m;p_1,p_2,p_3)&=\mu	\langle 1, v-y \log v\rangle+\lambda\| P_w(x)\|_0+\| z\|_l+\delta_{\mathbb{R}^n_+}(m)\nonumber\\
	&+\frac{\gamma}{2}\|kx-v+\gamma^{-1}p_1\|^2_2
	+\frac{\eta}{2}\|x-m+\eta^{-1}p_2\|^2_2\nonumber\\
	&+\frac{\beta}{2}\|\nabla x-z+\beta^{-1}p_3\|^2_2, 
\end{align}
where $\gamma$, $\eta$, $\beta$ are positive penalty parameters and 
$p_1,p_2$, $p_3$ are the Lagrangian multipliers.  
The iterative scheme based on ADMM is given as
\begin{align}\label{sp1}
v^{j+1}=\arg\min_{v > 0}L(x^j,v,z^j,m^j;p^j_1,p^j_2,p^j_3)=\arg\min_{v} \mu	\langle 1, v-y \log v\rangle+\frac{\gamma}{2}\|kx^j-v+\gamma^{-1}p^j_1\|^2_2,
\end{align}
\begin{align}\label{sp2}
x^{j+1}&=\arg\min_{x}L(x,v^{j+1},z^j,m^j;p^j_1,p^j_2,p^j_3)=\arg\min_{x}\frac{\gamma}{2}\|kx-v^{j+1}+\gamma^{-1}p^j_1\|^2_2
	+\frac{\eta}{2}\|x-m^j\nonumber\\
	&+\eta^{-1}p^j_2\|^2_2
	+\lambda\| P_w(x^j)\|_0
	+\frac{\beta}{2}\|\nabla^{\alpha} x-z^j+\beta^{-1}p^j_3\|^2_2, 
\end{align}
\begin{align}\label{sp3}
&z^{j+1}=\arg\min_{z}L(x^{j+1},v^{j+1},z,m^j;p^j_1,p^j_2,p^j_3)=\arg\min_{z} \|z\|_l+\frac{\beta}{2}\|\nabla^{\alpha} x^{j+1}-z+\beta^{-1}p^j_3\|^2_2,\\\label{sp4}
&m^{j+1}=\arg\min_{m}L(x^{j+1},v^{j+1},z^{j+1},m;p^j_1,p^j_2,p^j_3)=\arg\min_{m} \delta_{\mathbb{R}^n_+}(m)+\frac{\eta}{2}\|x^{j+1}-m+\eta^{-1}p^j_2\|^2_2,\\\nonumber
&p^{j+1}_1=p^{j}_1+\gamma (kx^{j+1}-v^{j+1}),\\\nonumber
&p^{j+1}_2=p^{j}_2+\eta (x^{j+1}-m^{j+1}),\\\nonumber
&p^{j+1}_3=p^{j}_3+\beta (\nabla^{\alpha}x^{j+1}-z^{j+1}).
\end{align}
We note that proposed algorithm is based on the ADMM and the convergence
analysis of this method is studied in \cite{24,25,26}.
Also, in the numerical section, the convergence of 
the proposed algorithm is studied using numerical results.

\subsection{Solving subproblems}
In the previous subsection, the subproblems that are required for image deblurring are introduced. In this subsection, the methods of solving these subproblems are explained.
It should be noted that in the following, for simplicity, we avoid using the power symbols $j$ and $j+1$. In a similar way to the quadratic formula for the roots of the general quadratic equation for subproblem \eqref{sp1}, we get
\begin{align}\label{eqf1}
v=\frac{\gamma kx+p_1-\mu+\sqrt{(\mu-\gamma kx-p_1)^2+4\mu \gamma g}}{2\gamma}.
\end{align}
Finding the answer for \eqref{sp2} is not as simple as \eqref{sp1} and we need to design a special method. By introduced auxiliary variable $n$, \eqref{sp2} is written as
\begin{align}\label{nsp1}
(x,n)&=\arg\min_{x,n}\frac{\gamma}{2}\|kx-v+\gamma^{-1}p_1\|^2_2
+\frac{\eta}{2}\|x-m
+\eta^{-1}p_2\|^2_2
+\lambda\|n\|_0\nonumber\\
&+\frac{\beta}{2}\|\nabla^{\alpha} x-z+\beta^{-1}p_3\|^2_2
+\frac{\rho}{2}\|n-P_w(x)\|^2_2, 
\end{align}
The new problem \eqref{nsp1} is divided into two subproblems as
\begin{align}\label{nsp2}
&x=\arg\min_{x}\frac{\gamma}{2}\|kx-\xi_1\|^2_2
+\frac{\eta}{2}\|x-\xi_2\|^2_2
+\frac{\beta}{2}\|\nabla^{\alpha} x-\xi_3\|^2_2
+\frac{\rho}{2}\|n-P_w(x)\|^2_2, \\\label{nsp3}
&n=\arg\min_{n}\lambda\|n\|_0+\frac{\rho}{2}\|n-P_w(x)\|^2_2,
\end{align}
where
$
\xi_1:=v-\gamma^{-1}p_1,~~\xi_2:=m-\eta^{-1}p_2,~~
\xi_3:=z-\beta^{-1}p_3
$.
Now, we define $x^w_P:=P_w^T(P_w(x))=Wx\circ M$ and
$\widehat{x}^w_P:=(1-M)\circ Wx$.  Where $M$ is the binary matrix and defined as
\begin{align*}
M_{i,j}=\begin{cases} 
	1, &\text{if (i,j) be a minimal pixel in patch,}\\
	0, & \text{other wise}.
\end{cases}
\end{align*}
It is easy to see that
$x=W^T\big(x^w_P+\widehat{x}^w_P\big)$.
For simplicity in performing calculations, we define 
$x_p:=W^Tx^w_P$
and $\widehat{x}_p:=W^T\widehat{x}^w_P$, therefore $x=x_p+\widehat{x}_p$. 
Based on these definitions, subproblem \eqref{nsp2} can be   
written as the two following subproblems.
\begin{align}\label{nsp4}
&x_p=\arg\min_{x_p} \frac{\gamma}{2}\|kx_p-\xi_4\|^2_2+
\frac{\eta}{2}\|x_p-\xi_5\|^2_2+\frac{\beta}{2}
\| \nabla^{\alpha}x_p-\xi_6\|^2_2
+\frac{\rho}{2}\|x_p-\xi_7\|^2_2,\\\label{nsp5}
&\widehat{x}_p=\arg\min_{\widehat{x}_p} \frac{\gamma}{2}\|k\hat{x}_p-\xi_8\|^2_2+
\frac{\eta}{2}\|\widehat{x}_p-\xi_9\|^2_2+\frac{\beta}{2}
\| \nabla^{\alpha}\widehat{x}_p-\xi_{10}\|^2_2,
\end{align}
where $\xi_i,~i=4,\cdots,10$ are defined as
\begin{align*}
&\xi_4:=\xi_1-k\widehat{x}_p,~~\xi_5:=\xi_2-\hat{x}_p,
~~\xi_6:=\xi_3-\nabla^{\alpha}\widehat{x}_p,~~\xi_7:=W^TP^T_w(n),\\
&\xi_8=\xi_1-kx_p,~~\xi_9=\xi_2-x_p,~~\xi_{10}=\xi_3-\nabla^{\alpha} x_p.
\end{align*}
By using the optimal condition for subproblems \eqref{nsp4}-\eqref{nsp5}, we get
\begin{align*}
&\big(\gamma k^T k+(\eta+\rho) I+\beta (\nabla^{\alpha})^T\nabla^{\alpha} 
\big)x_p=\gamma k^T\xi_4+\eta \xi_5+\beta (\nabla^{\alpha})^T\xi_6+\rho \xi_7,\\
&\big(\gamma k^T k+\eta I+\beta (\nabla^{\alpha})^T\nabla^{\alpha} 
\big)\widehat{x}_p=\gamma k^T\xi_8+\eta \xi_9+\beta (\nabla^{\alpha})^T\xi_{10}.
\end{align*}
These subproblems can be easily solved by assuming the periodic boundary conditions and using fast Fourier transform. Therefore, $x_p$ and $\widehat{x}_p$ are obtained by 
\begin{align}\label{eqf2}
&x_p=F^{-1}\Big(\frac{\gamma F(k^T\xi_4)+\eta F(\xi_5)+\beta F\big((\nabla^{\alpha})^T\xi_6\big)+\rho F(\xi_7)}
{\gamma F^{*}(k) F(k)+\eta+\rho+\beta F^{*}(\nabla^{\alpha})F(\nabla^{\alpha})} \Big),\\\label{eqf3}
&\widehat{x}_p=F^{-1}\Big(\frac{\gamma F(k^T\xi_8)+\eta F(\xi_9)+\beta F\big((\nabla^{\alpha})^T\xi_{10}\big)}
{\gamma F^{*}(k) F(k)+\eta +\beta F^{*}(\nabla^{\alpha})F(\nabla^{\alpha})} \Big).
\end{align}
As a final step in approximating $x$, by using the proximal mapping of $l_0$-norm, the solution of the subproblem \eqref{nsp3}
is obtained as follows.
\begin{align}\label{Es14}
	n=\begin{cases} 
		P_w(x), &|P_w(x)|^2 \geq \frac{2\lambda}{\rho},\\
		0, & \text{other wise}.
	\end{cases}
\end{align}
In the following algorithm, the solution of the subproblem \eqref{sp3} for $l=1$ can be find by using 
the shrinkage operator as
\begin{align}\label{eqf4a}
z=\text{shrink}\big(\nabla^{\alpha}x+\beta^{-1},\beta^{-1} \big),
\end{align}
where $\text{shrink}(a,b)=\sign(a)\max\big(|a|-b,0\big)$. 
Also for $l=0$, the solution of the subproblem \eqref{sp3}
can written as 
\begin{align}\label{eqf4b}
	z=\begin{cases} 
	\nabla^{\alpha}x, &|\nabla^{\alpha}x|^2 \geq \frac{2}{\beta},\\
	0, & \text{other wise}.
\end{cases}
\end{align}
And finally,
$m$ in the subproblem \eqref{sp4} is obtained by using the following 
formula.
\begin{align}\label{eqf5}
m=\max\big(x+\eta^{-1}p_2,0\big).
\end{align}
The proposed algorithm, which is discussed in details in the above, 
is summarized as Algorithm \textcolor[rgb]{0.00,0.00,1}{1}.

\noindent\line(1,0){280}\\
\vspace{-0.3cm}
Algorithm 1: Non-blind image restoration algorithm.\\
\vspace{-0.3cm}
\noindent\line(1,0){280}\\
\begin{algorithmic}
	\State \textbf{Input:} Blurred image $y$, blur kernel $k$.
	\State \textbf{Choose:}  $\mu,\lambda,\gamma,\beta,\rho$.
	\State \textbf{Initialize:} $x^{0}=y,n^{0}=W(y),z=D(y),$
	\State $~~~~~~~~~~~~~~~~~~~~~~m^{0}=z^{0}=p^{0}_1=p^{0}_2=p^{0}_3=b^{0}=0$.
	\State $j=0$
	\Repeat
	\State obtain $v^{j+1}$ by \eqref{eqf1}.
	\State obtain $x_p$ by \eqref{eqf2}.
	\State obtain $\widehat{x}_p$ by \eqref{eqf3}.
	\State consider $x^{j+1}=x_p+\widehat{x}_p$.
	\State obtain $n$ by \eqref{Es14}.
	\State obtain $z^{j+1}$ by \eqref{eqf4a} for $l_1$-FPMP or \eqref{eqf4b} for $l_0$-FPMP.
	\State obtain $m^{j+1}$ by \eqref{eqf5}.
	\State update $p^{j+1}_1,p^{j+1}_2$ and $p^{j+1}_3$ by
	\State$~~~p^{j+1}_1=p^{j}_1+\gamma (kx^{j+1}-v^{j+1}),$
	\State$~~~p^{j+1}_2=p^{j}_2+\eta (x^{j+1}-m^{j+1}),$
	\State$~~~p^{j+1}_3=p^{j}_3+\beta (\nabla^{\alpha}x^{j+1}-z^{j+1})$.
	\State $j=j+1$.
	\Until{Stop condition is met}
	\State \textbf{Output:} Deblurred image $x$.
\end{algorithmic}
\line(1,0){280}\\
\subsection{Blind deconvolution problem}
In the previous subsections, non-blind problem is studied.
But in the real world, sometimes we have no information about 
the PSF, and we are faced with the non-blind type of problem.
In this subsection, we deal with this type of problem. The proposed 
model for this type is as follows:
\begin{align*}
	\min_{x \geq 0,k\in \omega} \mu\langle 1,kx-y \log(kx)\rangle+\lambda\| P_w(x)\|_0+\| \nabla^{\alpha} x\|_1+\varrho\| \nabla k\|_1
+\delta_{\mathbb{R}^N_+}(x),  
\end{align*}
where $\omega:=\{k\in \mathbb{R}^{l\times s}| k_{i,j}\geq 0, \sum_{i,j}k_{i,j}=1\}$.
This problem is divided into two sub-problems. 
In the first problem, the original image is approximated,
and in the second problem, the blurred kernel is approximated.
The first problem is studied in the previous subsection and the second problem can be expressed as follows
\begin{align*}
	\min_{k\in \omega} \mu\langle 1, kx-y \log(kx)\rangle+\varrho \| \nabla k\|_1, 
\end{align*}
In $(j+ 1)$th iterative procedure, the solution of this problem is obtained
as (see \cite{27})
\begin{align}\label{aw0}
&k_t^{j+1}=\frac{k^{n}}{1-\frac{\varrho}{\mu} div\big(\frac{\nabla k^{j}}{|\nabla k^{j}|}\big)}
\big( X^{j} \big)^T\Big[\frac{y}{X^{j}k^{j}}\Big],\\\label{aw1}
&k^{j+1}=\frac{k_t^{j+1}}{\sum k_t^{j+1}}.
\end{align}
It should be noted that as mentioned in the introduction section to solve
ringing artifacts near the image borders, 
the method based on
Fourier domain restoration filter and extrapolated image
\cite{18} is used in this algorithm.
The proposed blind deblurring algorithm 
is summarized in Algorithm \textcolor[rgb]{0.00,0.00,1}{2}.

\noindent\line(1,0){280}\\
\vspace{-0.3cm}
Algorithm 2: Blind image restoration algorithm.\\
\vspace{-0.3cm}
\noindent\line(1,0){280}\\
\begin{algorithmic}
	\State \textbf{Input:} Blurred image $y$.
	\State \textbf{Choose:}  $\mu,\lambda,\gamma,\beta,\rho,\varrho$.
	\State \textbf{Initialize:} $x^{0}=y,n^{0}=W(y),z=D(y),$
	\State $~~~~~~~~~~~~~~~~~~~~~~m^{0}=z^{0}=p^{0}_1=p^{0}_2=p^{0}_3=b^{0}=0$,
	\State $~~~~~~~~~~~~~~~~~~~~~$ Consider $k^0$ as average kernel.
	\State prepare the blurred image using algorithm in \cite{18}.
	\State $j=0$
	\Repeat
	\State update $x^{j+1}$ by Algorithm \textcolor[rgb]{0.00,0.00,1}{1}.
	\State update $k^{j+1}$ by \eqref{aw0}-\eqref{aw1}
	\State $j=j+1$.
	\Until{Stop condition is met}
	\State \textbf{Output:} Deblurred image and kernel.
\end{algorithmic}
\line(1,0){280}\\

\section{Experiment results}\label{sec4}
In the previous section, a model for Poissonian blurred image deconvolution is introduced and then a numerical algorithm is proposed to solve this model. In this section, in order to show the efficiency of the algorithm, some examples are studied.
To simulate the algorithm, the system based on
Windows 10-64bit and Intel(R) Core(TM) i3-5005U CPU$@$2.00GHz,
and MATLAB 2014b are used.
In the simulation part, \verb"fspecial" and \verb"poissrnd"
are used to generate PSF and Poissonian noise, respectively.
Also different values are used in the following examples for the peak to check the noise intensity.
In the numerical results by using Frobenius norm, the stop condition is considered as
$Error=\|x^{j+1}-x_{j}\|_F/\|x^{j}\|_F \leq tol$, where $10^{-4}$ and $10^{-3}$ are chosen as $tol$ for nonblind and blind
problems, respectively. 
In this section to check the algorithm, the 
peak signal to noise ratio (PSNR) 
and the structural similarity (SSIM)
values are calculated.
The differences and similarities between
the original and restored image is measured by PSNR
an obtained as
\begin{align*}
	PSNR=10\log10\frac{p^2}{MSE(u,U)},
\end{align*}
where $u$ and $U$ denote original and restored image, respectively,
$p$ shows the maximum pixel value and $MSE$ is mean squared error.
Also $SSIM$ is calculated by
\begin{align*}
	SSIM=\frac{(\mu_o\mu_r+C_1)(2\sigma_{or}+C_2)}{(\mu^2_o+\mu^2_r+C_1)(\sigma_o+\sigma_r+C_2)},
\end{align*}
where $\mu_i, \sigma_i,~i=o,r$ are the means and variances of the clear and deblurred images, respectively,
and $\sigma_{or}$ shows covariance between the clear and deblurred images.
The following values are used in the simulation section and the best output are selected after calculating.
patch size$\in \{ 12,15,25,35\}$,
$\alpha \in \{0.1 i| i=1,\cdots,10\}$,
$\lambda, \rho, \eta \in \{0.25,0.5,1,2,4\}$,
$\beta, \gamma \in \{ 10e-i|i=1,2,3,4,5\}$,
$\mu \in \{5,100,200,300\}$ and 
$\varrho/\mu \in \{0.1, 1, 5 ,20\}$.

\subsection{Non-blind problem}
\begin{figure}[H]
	\centering
	\makebox[0pt]{
	\subfigure[clear image($I$)]{\label{fig:gull}\includegraphics[width=1\textwidth]{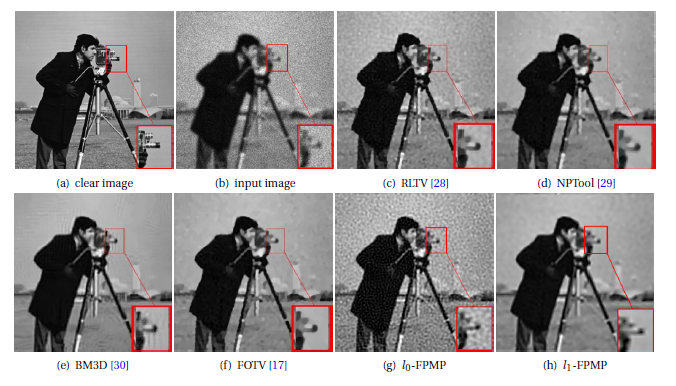}}}\\
	\vspace{-.2cm}
	\caption{Visual comparison of Cameraman with peak $255$.}
	\label{fig2}
\end{figure}

\begin{figure}[H]
	\centering
	\makebox[0pt]{
	\subfigure[clear image($I$)]{\label{fig:gull}\includegraphics[width=1\textwidth]{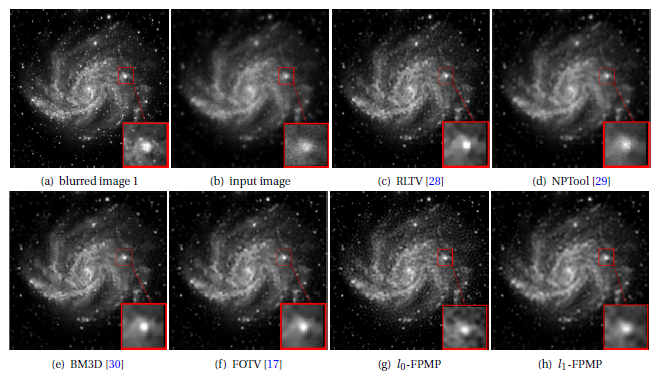}}}\\
	\vspace{-.2cm}
	\caption{Visual comparison for Galaxy with peak $255$.}
	\label{fig3}
\end{figure}

\noindent For the first example, the image of Cameraman with peak $255$ is blurred by
Gaussian PSF with size $9$ and standard deviation $\sqrt{3}$. The results of the proposed algorithm
is shown and compared in Fig. \ref{fig2}.
By using results in this figure, it is observed that the use $l_1$ norm compared to $l_0$ norm makes the restored image smoother. Also, for a more accurate comparison, the value of the PSNR is given in Table \ref{tab1} for two images, Cameraman and Galaxy. In these cases, the previous PSF is considered.
According to these results, as seen in Fig. \ref{fig2}, the use of $l_0$ norm does not always create a suitable answer, but the use of $l_1$ norm causes a more appropriate result with the compared algorithms.
In order to examine the astronomical image, the results of the proposed algorithm for Galaxy in Fig. \ref{fig3} are compared with other algorithms. Based on the simulation results, we can see that the proposed method compares
favorably or even better against compared algorithms. 

\begin{figure}[H]
	\centering
		\makebox[0pt]{
		\subfigure[clear image($I$)]{\label{fig:gull}\includegraphics[width=1\textwidth]{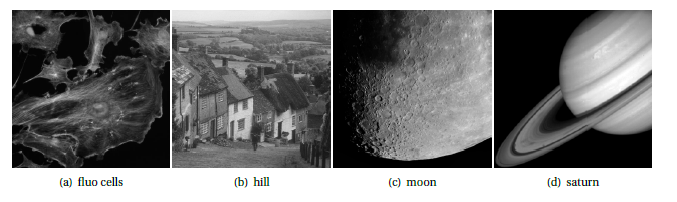}}}\\
	\vspace{-.2cm}
	\caption{Test images.}
	\label{fig4}
\end{figure}

\begin{figure}[H]
	\centering
		\makebox[0pt]{
		\subfigure[clear image($I$)]{\label{fig:gull}\includegraphics[width=1\textwidth]{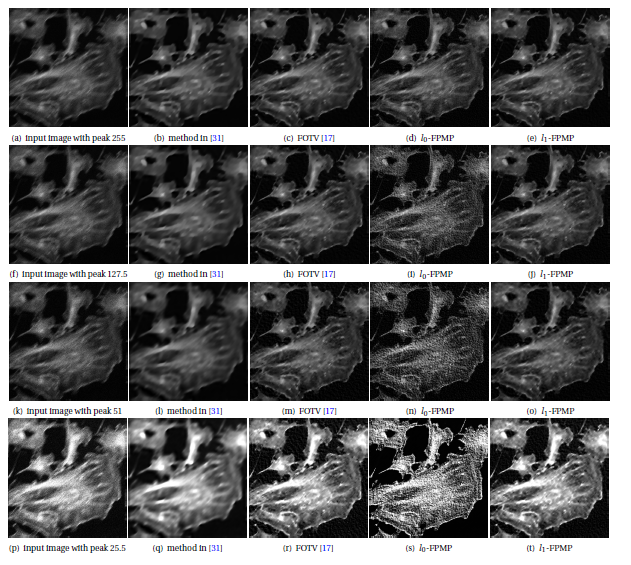}}}\\
	\vspace{-.2cm}
	\caption{Visual comparison for fluo cells image with different peaks.}
	\label{fig5}
\end{figure}

\begin{figure}[H]
	\centering
		\makebox[0pt]{
		\subfigure[clear image($I$)]{\label{fig:gull}\includegraphics[width=1\textwidth]{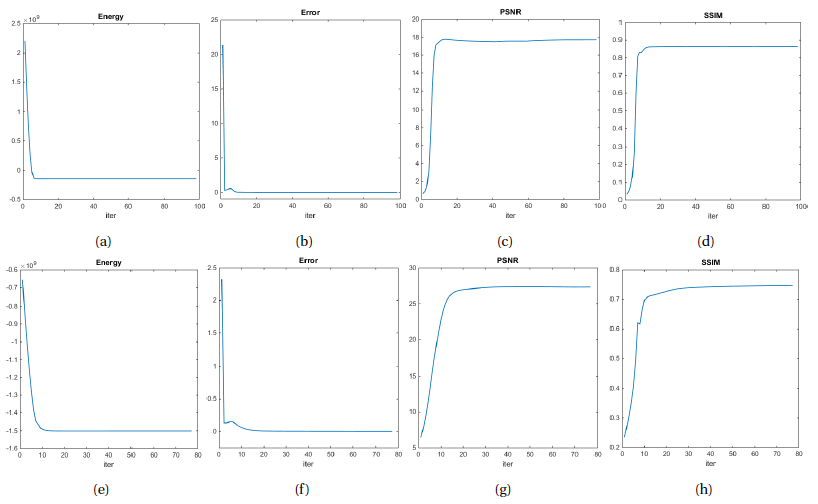}}}\\
	\vspace{-.2cm}
	\caption{Energy, PSNR, Error and SSIM curves for fluo cells image with peak$=255$, (a-d) $l_0$-FPMP, (e-h) $l_1$-FPMP.}
	\label{fig6}
\end{figure}

\begin{center}
\begin{table}[H]
\footnotesize
\caption{{\footnotesize Compared results of	PSNR for different
		images and peaks.}}
\label{tab1}
\centering
\begin{tabular}{ccccccccccccccc}
\hline
Imgae       &Peak  &RLTV \cite{28} &NPTool \cite{29} &BM3D \cite{30} &FOTV \cite{17}  &$l_0$-FPMP  &$l_1$-FPMP\\
\hline
	    	&255   &24.49 &23.74  &24.56 &24.59   &24.33 &25.62 \\
Cameraman   &127.5 &23.87 &23.90  &24.20 &24.14   &23.78 &25.72\\
			&51    &22.55 &23.21  &23.58 &23.40   &21.24&24.32\\
			&25.5  &20.96 &22.49  &23.09 &22.75   &19.56 &23.67\\
			\hline
			&255   &25.57 &24.64  &25.04  &25.34  &25.73 &26.65\\
Galaxy      &127.5 &24.99 &24.28  &24.66 &24.90&25.43 &26.07\\
			&51    &24.24 &24.12  &24.14  &24.38  &24.82 &25.54\\
			&25.5  &23.28 &23.83  &23.77  &23.86  &22.28 &24.41\\
					\hline
		\end{tabular}
	\end{table}
\end{center}
In the following, in order to investigate the algorithm, we consider four images, including 
fluo cells, hill, moon and saturn images. These images are shown in Fig. \ref{fig4}.
For fluo cells and hill motion PSF with 20 pixels and an angle of 45 degree in a counterclockwise direction
is used. Also Gaussian PSF with size $9$ and standard deviation $\sqrt{3}$ is used for moon and saturn images.
The results of PSNR and SSIM values are given in Table \ref{tab2}.
Numerical results are given in this table for different values of peak and compared with other algorithms.
In this table, as seen in the previous examples, the results of using $l_0$ norm have less accuracy than the algorithm based on $l_1$ norm. The results for the restored images for different peaks for fluo cells image are given in Fig \ref{fig5}. These results are compared with other algorithms. The results show the efficiency of the proposed method.
In order to investigate the convergence of the proposed algorithm, the figures of 
Energy, PSNR, Error and SSIM curves for fluo cells image with peak$=255$ are drawing in Fig. \ref{fig6}.
As it is clear from these results, after a number of repetitions, the resulting value approaches a certain value.

\begin{center}
	\begin{table}[H]
			\footnotesize
			\caption{{\footnotesize Compared results  of PSNR and SSIM for different images and peaks.}}
			\label{tab2}
			\centering
			\begin{tabular}{ccccccccccccccc}
				\hline
				 &&\multicolumn{2}{c}{method in \cite{31}}   &\multicolumn{2}{c}{FOTV \cite{17} }   &\multicolumn{2}{c}{$l_0$-FPMP } &\multicolumn{2}{c}{$l_1$-FPMP}\\\cmidrule(l){3-4} \cmidrule(l){5-6} \cmidrule(l){7-8} \cmidrule(l){9-10}
				Image&Peak   &PSNR &SSIM&PSNR &SSIM &PSNR &SSIM &PSNR &SSIM\\
				\hline
				&255    &24.92 &0.69    &26.84 &0.74  &26.16  &0.61&27.28 &0.75\\
	fluo cells	&127.5  &24.11 &0.81    &26.98 &0.84  &22.73  &0.59&27.26 &0.84\\
				&51     &22.48 &0.92    &26.41 &0.93  &21.98  &0.78&26.78 &0.94\\
				&25.5   &16.55 &0.82    &18.04 &0.82  &18.14  &0.59&18.18 &0.83\\
				\hline
				&255    &22.38 &0.45    &22.67 &0.45  &22.14 &0.45&23.21 &0.47\\
	hill    	&127.5  &22.04 &0.63    &23.03 &0.64  &22.10 &0.48&23.40 &0.65\\
				&51     &21.40 &0.85    &22.88 &0.86  &20.72 &0.69&22.94 &0.85\\
				&25.5   &21.11 &0.94    &22.81 &0.94  &17.51 &0.49&22.99 &0.94\\
				\hline
				&255    &25.02 &0.67    &25.65 &0.70  &26.68 &0.72&27.70 &0.73\\
	moon    	&127.5  &24.33 &0.79    &25.63 &0.82  &25.08 &0.69&26.19 &0.82\\
				&51     &23.55 &0.91    &25.57 &0.93  &23.16 &0.83&25.97 &0.93\\
				&25.5   &23.36 &0.96    &25.16 &0.96  &20.26 &0.63&25.47 &0.97\\
				\hline
				&255    &32.26 &0.89  &31.24 &0.90  &31.83&0.90&33.34&0.91\\
	saturn    	&127.5  &31.12 &0.93  &30.94 &0.93  &31.37&0.93&31.80&0.94\\
				&51     &29.64 &0.96  &30.44 &0.97  &30.11&0.96&31.14&0.97\\
				&25.5   &28.42 &0.97  &29.53 &0.98  &27.89&0.96&30.27&0.98\\
				\hline
		\end{tabular}
		\end{table}
\end{center}

\subsection{Blind problem}
\begin{figure}[H]
	\centering
	\makebox[0pt]{
	\subfigure[clear image($I$)]{\label{fig:gull}\includegraphics[width=1\textwidth]{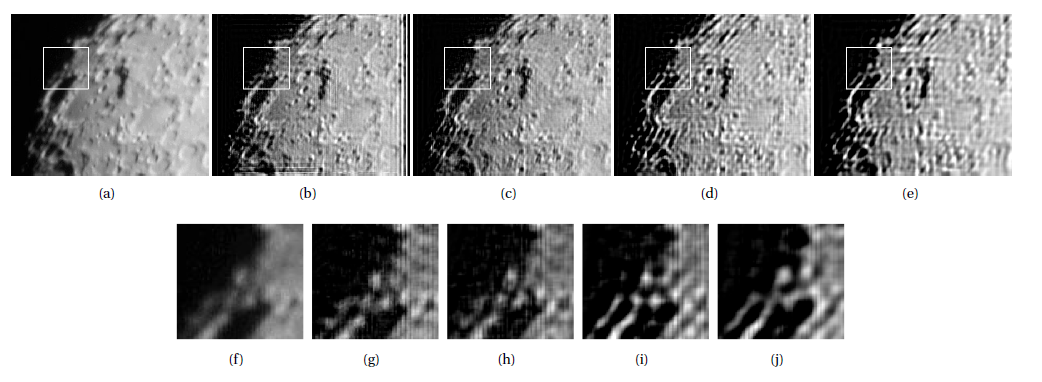}}}\\
	\vspace{-.2cm}
	\caption{Visual results for real image.}
	\label{fig8}
\end{figure}

\begin{figure}[H]
	\centering
		\makebox[0pt]{
		\subfigure[clear image($I$)]{\label{fig:gull}\includegraphics[width=1\textwidth]{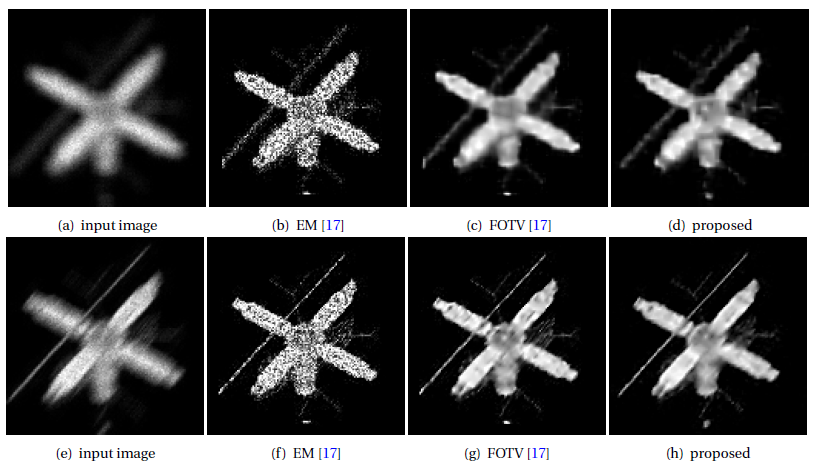}}}\\
	\vspace{-.2cm}
	\caption{Visual comparison for satellite images.}
	\label{fig7}
\end{figure}

\noindent Now the proposed algorithm for blind case is studied in this part.
As first example, the spin image is blurred by PSF $k_1$ and $k_2$
where $k_1$ is consider as Gaussian PSF with size $9$ and standard deviation $20$
and $k_2$ is obtained as motion PSF with 11 pixels and an angle of 50 degree.
The results of proposed algorithm for PSNR value are compared in Table \ref{tab3} for different kernels and peaks.
Also, the results related to restored images are given in Fig.\ref{fig7} and compared
These results indicate the efficiency of the proposed method in image restoring.
In the last example of this section, we study the real image using the proposed algorithm.
For this example, different values is chosen for the kernel size as $9\times 9$ (Fig.\ref{fig8}-(b,c)), $12\times 12$ (Fig.\ref{fig8}-(d))
and $15 \times 15$ (Fig.\ref{fig8}-(e)).
The results are shown in Fig. \ref{fig8}.
Comparing the results with the original image, it can be seen that the restored images 
obtained from the proposed algorithm reveal more details about the surface of the moon 
that is hidden in the original image.
In these figures, Fig.\ref{fig8}-(b,g) show the restored image without using algorithm in \cite{18}.
Comparing this result with the images in which the algorithm \cite{18} is used, it is observed that the ringing artifacts is created near the image borders.
\begin{center}
\begin{table}[H]
	\footnotesize
	\caption{{\footnotesize  Compared results  of PSNR for satellite image.}}
	\label{tab3}
	\centering
	\begin{tabular}{ccccccccccccccc}
		\hline
		Imgae   &Peak   & &EM \cite{17} &&FOTV \cite{17}   & &proposed       \\
		\hline
		$k_1$   &255    & &14.92 &&21.44   &&22.18\\
	        	&127.5   &&14.72 &&21.26   &&22.56 \\
		\hline
		$k_2$  &255      &&16.80 &&23.44 &&24.56\\
		       &127.5   & &16.63 &&20.98  &&24.36  \\
		\hline
	\end{tabular}
\end{table}
\end{center}

\section{Conclusion}\label{sec5}
In this paper the nonblind and blind Poissonian blurred image deconvolution 
algorithm  based on the local minimal prior are introduced.
In the local minimal prior, Framelet transform is used to find the difference between the clear and the blurred images. Also, the alternating direction method of multipliers (ADMM) is used to solve the proposed model.
In the blind deconvolution algorithm, Fourier domain restoration filter
is used to remove the ringing artifacts in the restored image.
The results of simulation algorithms have been compared with different methods and these results confirm the efficiency of the proposed algorithms.

\end{document}